\documentclass[3p,times]{elsarticle}

\usepackage{ecrc}

\usepackage{epstopdf}

\volume{00}

\firstpage{1}

\journalname{Nuclear Physics A}

\runauth{K.~C.~Zapp and S.~Floerchinger}

\jid{nupha}

\jnltitlelogo{Nuclear Physics A}

\usepackage{graphicx}
\usepackage{amsmath,amssymb}
\usepackage{units}

\begin{document}

\begin{frontmatter}



\title{Interplay between hydrodynamics and jets}

\author{Korinna C. Zapp}
\author{Stefan Floerchinger}

\address{Department of Physics, Theory Unit, CERN, CH-1211 Geneva 23}

\begin{abstract}
By combining the jet quenching Monte Carlo \textsc{Jewel} with a realistic hydrodynamic model for the background we investigate the sensitivity of jet observables to details of the medium model and quantify the influence of the energy and momentum lost by jets on the background evolution. On the level of event averaged source terms the effects are small and are caused mainly by the momentum transfer.
\end{abstract}


\end{frontmatter}

\section{Introduction}

While properties of soft particles produced in heavy-ion collisions can be understood in a hydrodynamic framework, it is clear that hard particles cannot be in local thermal equilibrium. The interplay between the soft, strongly coupled and the hard, weakly coupled sector gives access to non-trivial QCD dynamics. Different techniques are used to describe these two regimes and there is currently no satisfactory approach allowing for a fully self-consistent description of the entire dynamics in a common framework. Instead, it is common practice to use perturbative techniques for hard probes and hydrodynamics for the soft bulk of the event. The drawback of this approach is that the separation between the soft and the hard regime is to some extent arbitrary and that it is difficult to fully account for the crosstalk between the two regimes. So far the modification of hard probes, particularly jets, due to interactions in the soft background and the resulting jet quenching phenomenology has received more attention than the modification of the bulk evolution. In a recent study~\cite{Floerchinger:2014yqa} we combined the jet quenching Monte Carlo \textsc{Jewel} with a Bjorken boost invariant and azimuthally symmetric  viscous hydrodynamic evolution of the bulk to get a realistic estimate of the effect that the passage of jets can have on the soft event.

\section{Combining hydrodynamics and jets}

On the hydrodynamic side viscous corrections are taken into account in a second order formalism including shear viscosity and the corresponding relaxation time, which take their AdS/CFT values (bulk viscosity is neglected). The equation of state is a parametrisation of a lattice equation of state combined with a hadron resonance gas (s95p-PCE  of~\cite{Shen:2010uy}). The system is assumed to be boost-invariant along the beam axis and to have azimuthal symmetry (corresponding to the most central $b=0$ collisions). The initial conditions are specified at $\tau = \unit[0.6]{fm}$ following~\cite{Qiu:2011hf}, i.e.\ $T=485\, \text{MeV}$ in the centre of the collision with a profile in the transverse plane given by a Glauber calculation, $u^r=0$ and the Navier-Stokes values of the shear stress.

\smallskip

The QCD evolution of jets in the presence of re-scattering in a dense background is simulated with \textsc{Jewel}~\cite{Zapp:2012ak}. The jet production points are distributed according to the density of binary nucleon-nucleon collisions extracted from a Glauber model~\cite{Eskola:1988yh}. The number of di-jets per event is Poisson distributed. The jet production matrix elements and initial state parton showers are generated with \textsc{Pythia}\,6~\cite{Sjostrand:2006za} using the EPS09 nuclear pdf sets~\cite{Eskola:2009uj}. In the absence of a medium the final state jet evolution reduces to a standard virtuality ordered parton shower similar to the one in \textsc{Pythia}\,6. In the presence of a background medium re-scattering can occur, that can be either elastic or inelastic, i.e.\ give rise to QCD bremsstrahlung. Re-scattering is described using leading order perturbative $2\to 2$ scattering matrix elements with radiative corrections being generated by the parton shower. The space-time structure of the parton shower and the interplay between different sources of radiation as well as the destructive LPM-interference are dictated by the formation times of the emissions.

\textsc{Jewel} has to be provided with information about the background, namely the local density and the momentum distribution of scattering centres. When running with the hydrodynamic medium it takes the local temperature and fluid velocity as input and constructs the parton density and momentum distribution from it assuming an ideal gas equation of state.

\smallskip

As \textsc{Jewel} is a completely microscopic model it is straightforward to extract the energy-momentum transfer between the jets and the medium in the individual scattering processes, 
\begin{equation}
J^\mu(x) = \sum_{i} \Delta p^\mu_i \delta^{(4)}(x-x_i) \,.
\end{equation}
This can interpreted as a source term in the hydrodynamic equations. It is convenient to decompose it into components parallel and orthogonal to the fluid velocity $u$,
\begin{equation}
J_S = u_\nu J^\nu \qquad \text{and} \qquad J_V^\mu = \Delta^\mu_{\;\;\nu} J^\nu \,,
\end{equation}
where $\Delta^{\mu\nu} = u^\mu u^\nu + g^{\mu\nu}$. The source term varies from event to event. One possible solution is to solve the hydrodynamic equations event-by-event, which is, however, computationally expensive. We therefore chose to characterise the statistical properties of the source term in terms of n-point functions. Assuming the fluctuations to be Gaussian it is sufficient to specify the event averages
\begin{equation}
 \bar J_S= \langle J_S(x) \rangle \qquad \text{and} \qquad \bar J_V^\mu =  \langle J_V^\mu(x) \rangle
\end{equation}
and the correlation functions 
\begin{equation}
\begin{split}
\bar C_{SS}(x,y) & = \langle J_S(x)J_S(y)\rangle - J_S(x) J_S(y) \\
\bar C_{SV}^\mu(x,y) & = \langle J_S(x)J_V^\mu(y)\rangle - J_S(x) J_V^\mu(y) \\
\bar C_{VV}^{\mu\nu}(x,y) & = \langle J_V^\mu(x)J_V^\nu(y)\rangle - J_V^\mu(x) J_V^\nu(y) \,.
\end{split}
\end{equation}
Then the fluid dynamic equations containing the average source term read
\begin{equation}
\begin{split}
& u^\tau \partial_\tau \epsilon + u^r \partial_r \epsilon + (\epsilon+p) (\partial_\tau u^\tau + \partial_r u^r + \tfrac{1}{\tau} u^\tau + \tfrac{1}{r} u^r)\\
& + u^\tau \left[ \partial_\tau \pi^{\tau\tau} + \tfrac{1}{\tau} \pi^{\tau\tau} + \partial_r \pi^{\tau r} + \tfrac{1}{r} \pi^{\tau r} + \tfrac{1}{\tau} \pi^{\eta\eta} \right]\\
& - u^r \left[ \partial_\tau \pi^{\tau r} + \tfrac{1}{\tau} \pi^{\tau r} + \partial_r \pi^{rr} + \tfrac{1}{r} \pi^{rr} - \tfrac{1}{r} \pi^{\phi\phi}\right] = - \bar J_S
\end{split}
\end{equation}
for the energy density $\epsilon$ and
\begin{equation}
\begin{split}
& (\epsilon + p) (u^\tau \partial_\tau u^r + u^r \partial_r u^r) + u^r u^\tau \partial_\tau (p+\pi_\text{bulk}) + (u^\tau)^2 \partial_r p \\
& - u^\tau u^r \left[ \partial_\tau \pi^{\tau\tau} + \tfrac{1}{\tau} \pi^{\tau\tau} + \partial_r \pi^{r\tau} + \tfrac{1}{r} \pi^{r\tau} + \tfrac{1}{\tau} \pi^{\eta\eta} \right] \\
& + (u^\tau)^2 \left[ \partial_\tau \pi^{\tau r} + \tfrac{1}{\tau} \pi^{\tau r} + \partial_r \pi^{rr} + \tfrac{1}{r} \pi^{rr} - \tfrac{1}{r} \pi^{\phi\phi}\right] = \bar J_V^r
\end{split}
\end{equation}
for the $\tau$-component of the fluid velocity ($\pi^{\mu\nu}$ denotes the shear stress tensor). The $\phi$ and $\eta$ component of the fluid velocity vanish due to the symmetries and $u^r$ is related to $u^\tau$ through the constraint $u_\mu u^\mu = -1$.

\section{Jets in the hydrodynamic background}

So far, \textsc{Jewel} was used mainly with a simple toy model as background, which assumes Bjorken expansion and an energy density that is proportional to the density of participants calculated in a Glauber model~\cite{Zapp:2012ak}. The initialisation time and initial temperature are the same in the toy model and the full fluid dynamical evolution, but the temperature profile in the transverse plane  is different (it falls off faster in the hydro initial conditions). Another important difference is that there is no transverse expansion in the toy model, so that the hydrodynamic background extends to larger radii at later times. Consequently, the average temperature in the toy model is higher at early times and lower later times than in the hydrodynamic solution.

When comparing jet observables computed with the two medium models but otherwise identical parameters one observes differences of at most \unit[20]{\%} in the jet nuclear modification factor (the differences in other observables are much smaller). The transverse expansion enters directly through the momentum distribution of the scattering centres and indirectly through the temperature distribution. The former has no visible effect on the final distributions while the latter is responsible for the observed differences. It turns out that the difference builds up only at late times ($\tau > \unit[4]{fm/c}$), when the temperature is already fairly small. The reason why there is sensitivity to the temperature distribution at late times in \textsc{Jewel} is that at early times the jet evolution is dominated by hard emissions associated to the jet production process and re-scatterings can only induce extra radiation at later times.

It should be noted that the differences between the two background models are smaller than other theory uncertainties even within the same jet quenching model.

\section{Results for event averaged source term}

When extracting the source terms a major difficulty, that is a consequence of the division into a soft and a hard regime, consists in defining the jet population. The perturbative jet cross section is infra-red divergent and has to be regularised. As we are interested in typical, i.e. minimum bias, events the jet cross section should cover the $p_\perp$ range where it dominates over the thermal background (partons with approximately thermal momenta will not contribute to the source terms as they on average don't lose energy and should be considered part of the background). This leads to a low $p_\perp$ cut on the jet production matrix element, $p_{\perp,\, \text{cut}} \simeq \unit[3]{GeV}$ and a correspondingly large number of di-jets per event ($N_\text{di-jet} \approx 1700$). The regularisation procedure also introduces a large uncertainty of the order of a factor 2 -- 3 on the source term. 

For central collisions the averaged source term will be independent of $\phi$, but it can have a non-trivial rapidity dependence even in a boost-invariant background. To preserve the symmetries of the bulk we consider in this study only the central rapidity unit (including the rapidity dependence is a straightforward extension).

\smallskip

\begin{figure}
\begin{center}
\includegraphics[angle=-90,width=.48\linewidth]{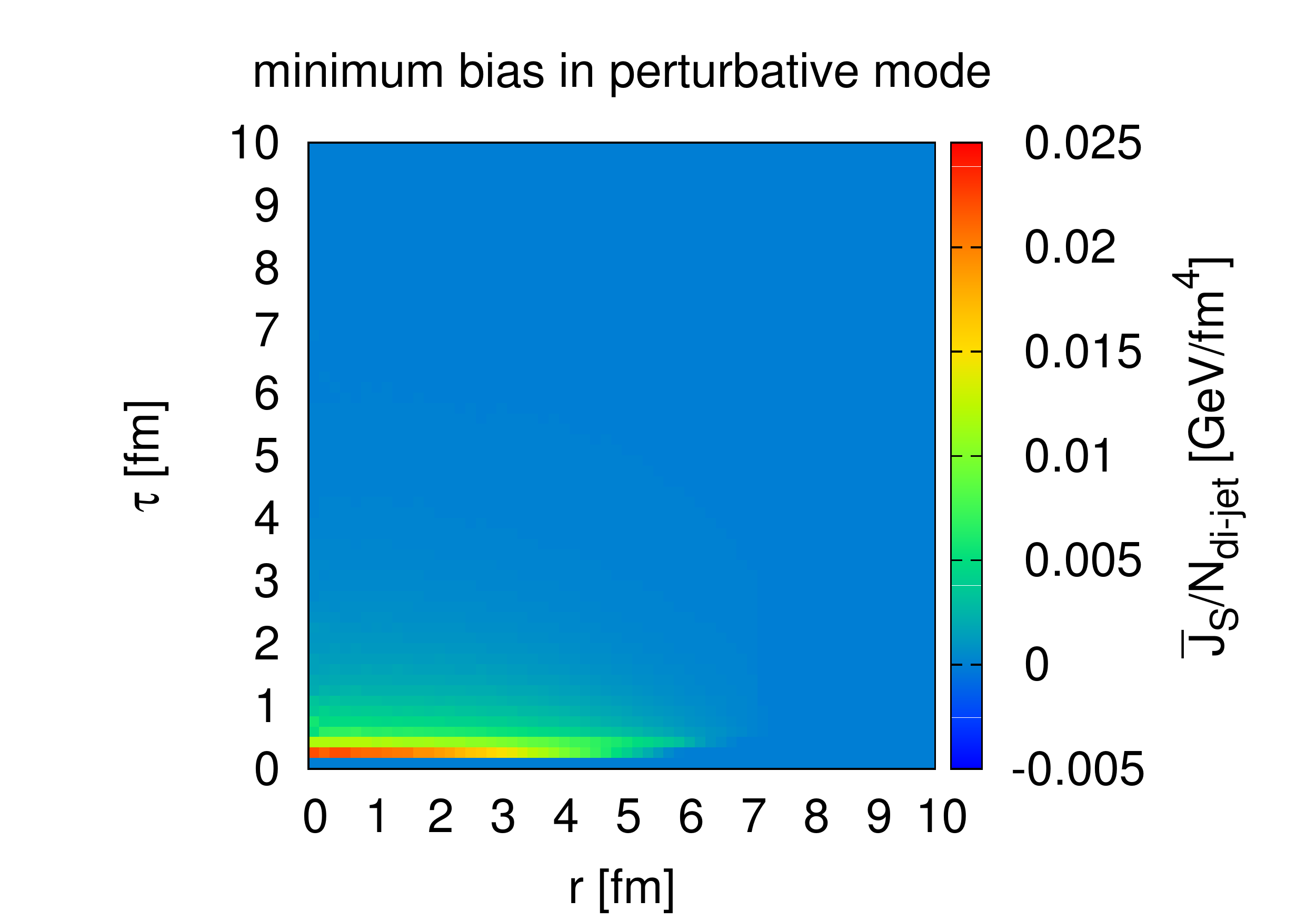}
\includegraphics[angle=-90,width=.48\linewidth]{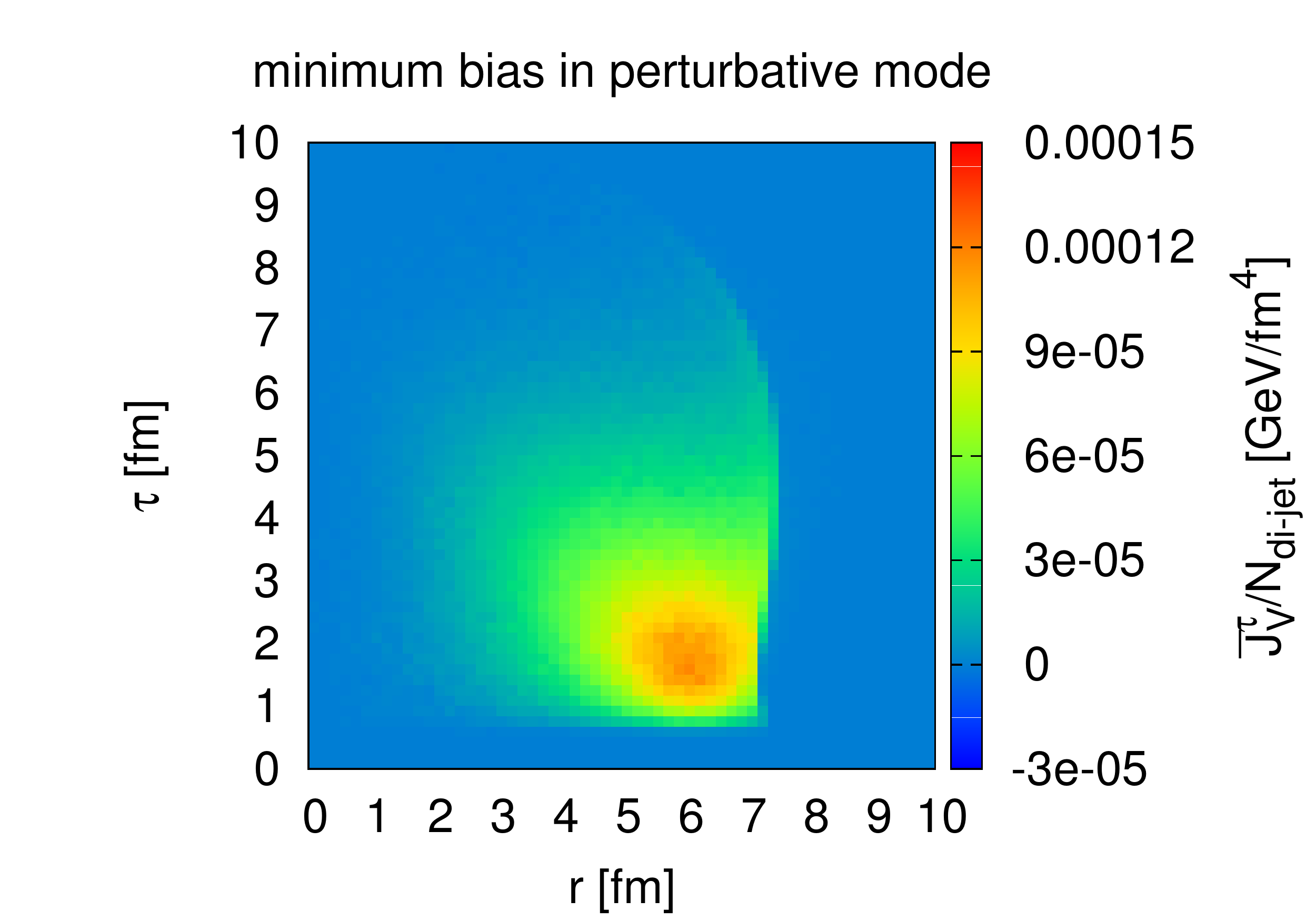}
\caption{Event averaged source terms $\bar J_S$ (left) and $\bar J_V^\tau$ (right) per di-jet in minimum bias events. The sources are averaged over the azimuthal angle $\phi$ and the central unit in rapidity.
}
\label{fig::sources}
\end{center}
\end{figure}

The averaged source terms scale trivially with the number of di-jets per event. In figure~\ref{fig::sources} the components $\bar J_S$ and $\bar J_V^\tau$ of the source term per event are shown ($\bar J_V^r$ is related to $\bar J_V^\tau$ through $u_\mu \bar J_V^\mu  =0$). The energy transfer $\bar J_S$ is concentrated at early times and small radii, i.e.\ where the temperature is highest. It falls off fast with $\tau$, as most 'jets' are soft 'mini-jets' that quickly evolve down to approximately thermal scales, where they do not contribute to the source terms any more. In contrast to this the momentum transfer builds up only at late times and at large radii. 

\smallskip

\begin{figure}
\begin{center}
\includegraphics[width=.45\linewidth]{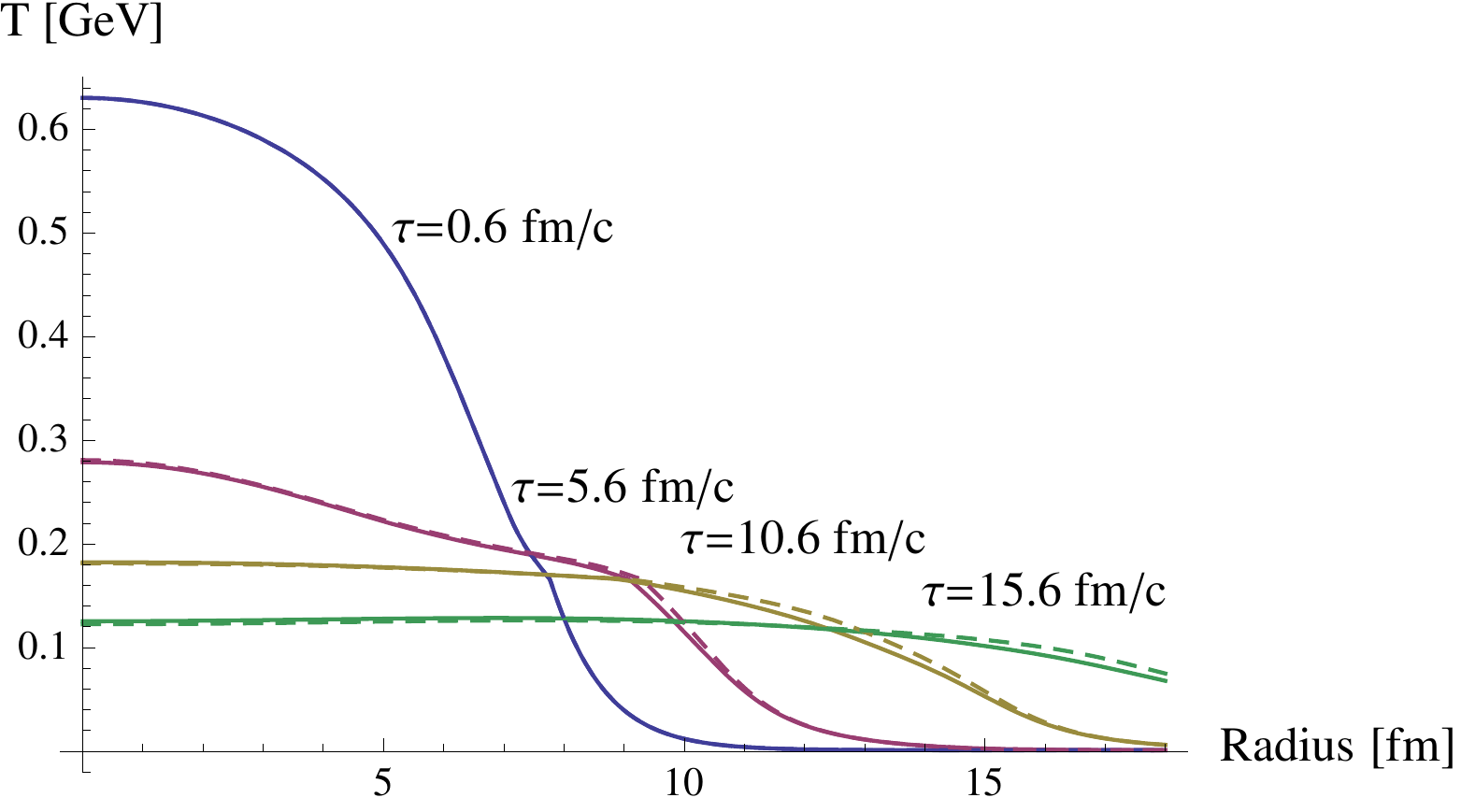}
\hspace{5mm}
\includegraphics[width=.45\linewidth]{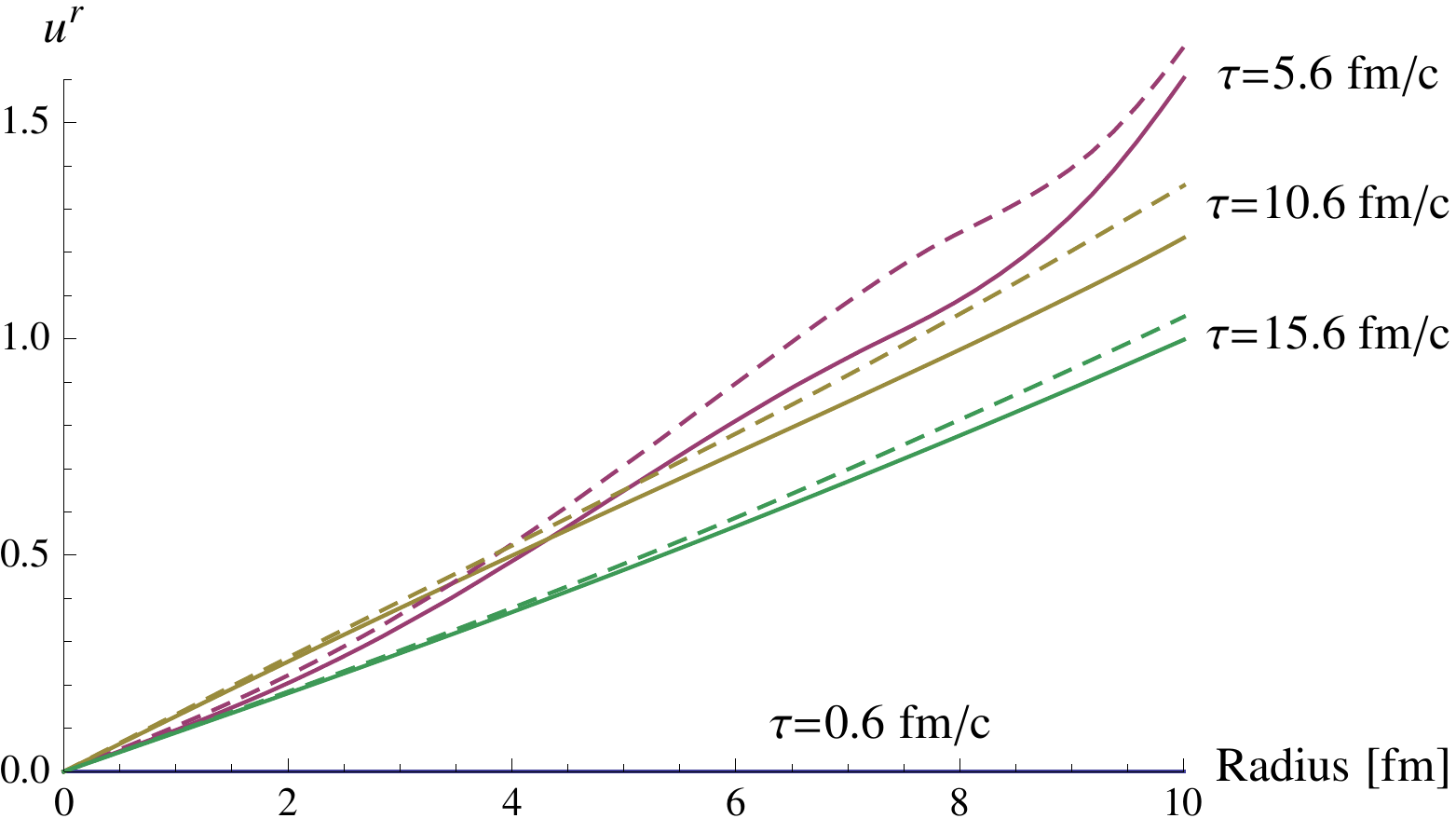}
\caption{Temperature (left) and radial component $u^r$ of the fluid velocity (right) for different times. Solid lines correspond to the solution without source terms, dashed lines show the solutions with the source terms shown in figure~\ref{fig::sources}.}
\label{fig::results}
\end{center}
\end{figure}

The effect of the averaged source term on the bulk evolution is shown in figure~\ref{fig::results}. It has only a very small influence on the temperature. The increase at early times is due to dissipation while the decrease in the centre and increase at large $r$ at later times is caused by the increase in the radial velocity. The latter is a consequence of the momentum transfer, which increases the transverse flow by up to \unit[10]{\%} at intermediate times.

\section{Conclusions}

We have studied the effect of a realistic fluid dynamic description of the background on jets and of energy and momentum loss by jets on the evolution of the background. On the jet side the differences between the hydrodynamic medium and a simple toy model were found to be smaller than other modelling uncertainties and caused by the fact that at later times the hydrodynamic background extends to larger radii due to the transverse expansion.

For the background evolution the energy and momentum deposited by jets forms source terms entering the fluid dynamic equations. We characterise the sources in terms of event averages and correlation functions and extract realistic estimates for minimum bias events. The averaged source terms are shown to have only a small influence on the radial flow caused by the momentum transfer, while the effect on the temperature is negligible. The influence of the correlation functions will be studied in a future publication.

The division of heavy ion events into a soft bulk described by hydrodynamics and hard jets described using perturbative techniques introduces sizeable uncertainties, as the separation between the two regimes is not well-defined. On the other hand, this approach allows one to use the best available descriptions for each of the two parts, a suitable common framework being currently unavailable. It is in this way also easier to separate different effects.

\end{document}